# On the reliable measurement of the gluon condensate in lattice QCD

Alexander Bochkarev [a]

Talk given at Lattice '94, Bielefeld, hep-lat/9411081

[a]Intitute of Theoretical Physics, University of Minnesota,
Minneapolis, MN 55455, USA

We propose to calculate the gluon condensate in lattice QCD in an indirect way by extracting it from the correlator of hadronic currents of heavy quarks. Moments (derivatives with respect to momentum at vanishing momentum) of the vector and pseudoscalar correlators are evaluated. The contribution of the continuum spectrum in addition to the low-lying resonance is observed in accordance with qualitative expectations. Finite size effects are observed and shown to be insignificant for heavy quarks and moderate lattices. A practical definition of the nonperturbatively renormalized heavy quark mass is given.

### Introduction

Extensive analysis of the two-point correlators of hadronic currents by means of the QCD sum rules indicates that vacuum expectation values of the quark and gluon fields receive large nonperturbative contributions (condensates) [1]. The existence of condensates implies a particular form of the crossover between the domain of asymptotic freedom and the domain of large distances, dominated by the low-lying resonance. There is certain range of intermediate virtualities where the perturbative series in powers of $\alpha_s$ is still under control, whereas nonperturbative effects show up, represented, in some channels, by the first few terms of the operator product expansion. It was recently suggested [2] that lattice simulations must be useful to study those intermediate distances in the physics of light quarks. We find it very advantageous to study deviations from the asymptotic freedom in the case of correlators of hadronic currents of heavy quarks. It is in this case that the gluon condensate was first introduced [1]. Such a way is an alternative to the direct evaluation of the gluon condensate as an average of the local operator, initiated in [3].

Consider the polarization operator of the electromagnetic currents of charmed quarks:

$$\Pi_{\mu\nu} = i \int dx\, e^{iqx} <0|T(j_\mu(x) j_\nu(0))|0>$$
$$\equiv \left(q_\mu q_\nu - q^2 g_{\mu\nu}\right) \Pi(q^2), \qquad (1)$$

where $j_\mu = \bar{c}\gamma_\mu c$. The function $\Pi(q^2)$ is computable in perturbative QCD at the origin $q^2 = 0$ since that point is far away from the leading threshold due to a pair of charmed quark and antiquark. $\Pi(q^2 = 0)$ is a function of the renormalized heavy quark mass $m_c(\mu)$ and $\alpha_s(\mu)$. The relevant renormalization point here $\mu = 2m_c$ is high enough to ensure the validity of the QCD perturbation theory: $\alpha_s(2m_c) \simeq 0.2$. Define moments of the polarization operator (1) as:

$$\mathcal{M}_n = \frac{1}{n!} \left\{ \left(\frac{d}{dq^2}\right)^n \Pi(q^2) \right\}_{q^2=0} \qquad (2)$$

Starting with $n = 2$ the moments $\mathcal{M}_n$'s are physical quantities: all the ultraviolet singularities are adsorbed by the bare Lagrangian parameters. Higher moments $n \gg 1$ correspond to smaller virtualities, which leads to deviations from the perturbative series before that perturbative series grows out of control [1]. For $n = 4 \div 7$ those deviations are well described by the few terms of the operator product expansion, starting with the quadratic gluon condensate $G^{(2)} \equiv <0|(\alpha_s/\pi) G^a_{\mu\nu} G^a_{\mu\nu}|0>$. For the ratios of neighboring moments $r_n = \mathcal{M}_n / \mathcal{M}_{n-1}$ one has on the two-loop level:

$$r_n = \frac{1}{4m_c^2}\left(a_n + b_n\, \alpha_s(4m_c^2) - c_n\, \frac{G^{(2)}}{(4m_c^2)^2}\right) \quad (3)$$

where $\{a_n, b_n, c_n\}$ are known numbers. The point we are making is that the l.h.s. of (3) is di-rectly accessible to the Monte-Carlo lattice simulations, and since the perturbative series in the r.h.s. of (3) is reliable the nonperturbative term represented by the gluon condensate is to be extracted in a reliable way. Lower moments ($r_{3,4}$) correspond to short distances and are dominated by the first two terms of (3), where $a_n$ corresponds to the one-loop contribution (asymptotic freedom term) and $b_n$ represents a small two-loop correction. Since the one-loop diagram is a function of only one parameter - the renormalized heavy quark mass, that parameter can be extracted from the first moments. In this way, using Monte-Carlo data on a finite lattice one can determine the nonperturbatively renormalized *heavy* quark mass in terms of the gauge invariant correlator without any reference to the pole mass. This mass is to be used later to extract the last term in the r.h.s. of (3) from the higher moments. For any given Monte-Carlo evaluation of the l.h.s. of (3) the perturbative part of the r.h.s. of that equation is to be built in terms of renormalised parameters of the improved lattice perturbation theory [4].

The contribution of the gluon condensate in (3) is about 10% relative to the leading first term $\sim a_n$ in the charmonium case. However, we are not forced to consider the phenomenological value of $m_c$. Adjusting the hopping parameter $\kappa$ so as to decrease $m_c$ from $1.3 GeV$ to, say $0.9 GeV$ where perturbative series in $\alpha_s(0.9 GeV)$ is still under control [1], one can easily increase the third term in (3) up to 20%. The efficiency of this trick is due to the fact that the perturbative terms of the r.h.s. of (3) depend on $m_c$ logarithmically, while the gluon condensate contribution is rapidly varying power correction $m_c^{-4}$.

On the lattice one has to deal with the original correlator $\Pi_{\mu\nu}(q^2)$ rather than $\Pi(q^2)$. We therefore introduce moments $\tilde{\mathcal{M}}_n$ as derivatives of $\Pi_{\mu\mu}(q^2)$ rather than $\Pi(q^2)$ in (2). The new moments are related to the initial ones (2) in a simple way $\tilde{\mathcal{M}}_n = -3\mathcal{M}_{n-1}$, so the ratio $\tilde{r}_n$ for the new moments will be $\tilde{r}_{n+1} = r_n$. To know the Fourier transform of the correlator (1) from the lattice simulations to the accuracy needed for the detection of the power corrections does not seem to be feasible. For this reason we evaluate not the Fourier transform $\Pi_{\mu\nu}(q^2)$ at nonzero $q^2$, but rather the moments themself directly in the $x$-space. At $q^2 = 0$ one can get rather simple expressions for the moments in terms of the original correlator $\Pi_{\mu\nu}(x)$:

$$\tilde{\mathcal{M}}_n = \frac{1}{2^{2n} n!(n+1)!} \int dx^4 \, x^{2n} \, \Pi_{\mu\mu}(x) \qquad (4)$$

**Monte-Carlo data**

We used the clover- and tadpole-improved quenched Wilson fermionic propagators of [5], obtained on the $8^3 \times 16$ lattice blocked from $32^3 \times 64$ $SU(3)$ configurations at $\beta = 6$. We have calculated the correlator (1) and the ratios of the moments (3) for vector (Fig.1) and pseudoscalar (Fig.2) currents ($j = i\bar{c}\gamma_5 c$ in (1)) for three values of the hopping parameter $\kappa = (0.9830, 0.1031, 0.1111)$. Statistical errors are noticeable but small for already 6 propagators ($\kappa = 0.1111$ - the upper curves) and negligible for 10 propagators ($\kappa = 0.0983$ - the lower curves). The given values of $\kappa$ correspond to *not* heavy quarks, so our data for $r_n$ should be compared to the QCD perturbation theory as in (3). We rather analyse the form of the spectral density $\rho(s)$ of the correlator (1) by looking at the functions $r(n)$.

The simple and phenomenologically very successful ansatz for $\rho(s)$:

$$\rho(s) = fm_{res}^2 \delta(s - m_{res}^2) + \frac{1}{4\pi^2}\theta(s - s_o) \qquad (5)$$

with a single low-lying resonance of mass $m_{res}$, residue $f$ and perturbative continuum with some effective threshold $s_o > m_{res}^2$ gives the ratios

$$\eta_n \equiv 1 + \frac{1}{4\pi^2 nf} \left(\frac{m_{res}^2}{s_o}\right)^n, \qquad (6)$$

$$r_n = \frac{1}{m_{res}^2} \frac{\eta_n}{\eta_{n-1}}$$

At large $n$ the ratios $r_n$ reach a constant $1/m_{res}^2$, which is the resonance contribution. That constant is approached from below due to the continuum contribution, which is just what one can see on Fig.1. As $n$ grows the ratios try to reach the asymptotic constant value from below, but at some point fail. That "crash"-point is a finite size effect. One cannot put arbitrarily high moments on a finite lattice because higher moments correspond to lower virtualities, larger distances. At some point they become too sensitive to the boundaries of the box. This interpretation is supported by the fact the "crash"-point is clearly $\kappa$-dependent. At greater $\kappa$, smaller $m_{res}$ (upper curves) the "crash" happens earlier. Figs. 1,2 demonstrate that one can easily accommodate heavy fermions ($\kappa < .0983$) on typically simulated lattices as far as the ratios of interest $r_2 \div r_7$ [1] are concerned.

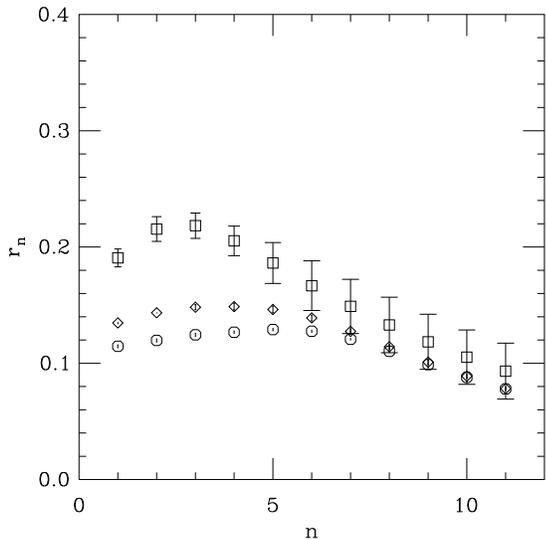

Figure 1. Ratios $r_n$ for three different hopping parameters in the vector channel.

The observed "asymptotic" values of the ratios do follow the law $\sim 1/m_{res}^2$. For instance, in the vector channel the height of the upper curve is 1.69 greater than the height of the lower curve. Whereas for the corresponding masses one has [5]:

$$\frac{m_{res}^2(\kappa = 0.0983)}{m_{res}^2(\kappa = 0.1111)} = 1.68 \qquad (7)$$

In the pseudoscalar channel (Fig.2), where the resonance (pion) mass is smaller, the ratios $r_n$ are flat near the origin, which means that the continuum is not seen. That is because the effective continuum threshold $s_o$ is significantly greater than resonance mass in that channel. For heavy quarks the continuum contribution will be more visible for small $n$ and allow for the comparison with the predictions of perturbative QCD. The Monte-Caro data with smaller lattice-spacings and smaller $\kappa$ are necessary to see the slope of the continuum contribution with good resolution at small $n$.

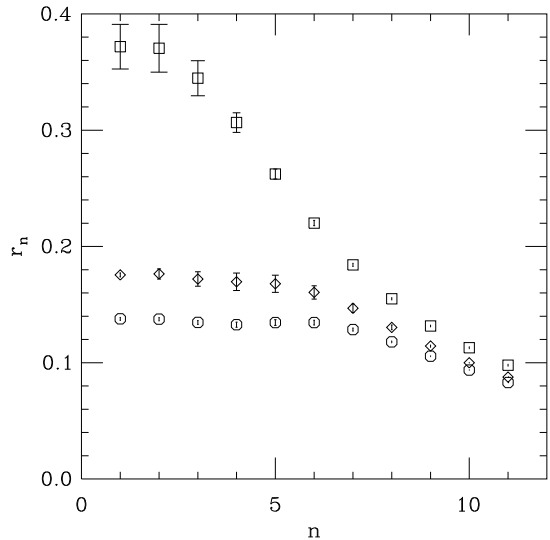

Figure 2. Same as in Fig. 1 in the pseudoscalar channel.

I am grateful to Ph. de Forcrand, L. McLerran, J. Smit, A. Vainshtein, M. Voloshin, R. Willey for useful discussions. This work was supported by the Pittsburgh Supercomuputer Center.